\documentclass[slac_one]{revtex4}
\usepackage{graphicx}
\usepackage{fancyhdr}
\usepackage{xspace}
\usepackage{floatflt}
\newcommand{\met}{\,/\!\!\!\!E_{T}}

\newcommand{\pairstop}{\tilde{t}_{1}\bar{\tilde{t}}_{1}}

\newcommand{\chargino}{\tilde{\chi}_{1}^{\pm}}
\newcommand{\neutralino}{\tilde{\chi}_{1}^{0}}

\newcommand{\ttbar}{t\bar{t}}

\begin{document}

\title{Search for Pair Production of Supersymmetric Top Quarks \\
Mimicking Standard Model Top Event Signatures at CDF}

%

\author{A. Ivanov (for the CDF Collaboration)}
\affiliation{University of California, Davis, CA, 95616, USA }

\begin{abstract}
We present results of the search for the super-symmetric partner of
the top quark, the stop quark $\left(\tilde{t}_{1}\right)$, decaying to a
$b$-quark and chargino $\left(\chargino\right)$ with the subsequent $\chargino$ 
decay  into a neutralino $\left(\neutralino\right)$, lepton $\left(\ell\right)$, and neutrino $\left(\nu\right)$. 
Using the data sample corresponding to 2.7~fb$^{-1}$ of integrated luminosity, 
collected with the CDF Detector of the Tevatron $p\bar{p}$ collider, we 
reconstruct the stop mass of candidate events and set 95\%~C.~L. upper limits
on masses of the stop quark, chargino and neutralino and the
branching ratio ${\cal B}(\chargino\rightarrow\neutralino\ell^{\pm}\nu)$. 
\end{abstract}

\maketitle

\thispagestyle{fancy}


\section{INTRODUCTION AND MOTIVATIONS} 

Supersymmetry (SUSY) is one of the most plausible extensions to 
the Standard Model (SM) of particle physics.
It naturally solves the problem with quadratically divergent 
quantum corrections contributing to the 
Higgs mass. It predicts unification of gauge coupling constants at a common GUT scale, and
provides a natural dark matter candidate. 
SUSY postulates that each of the fundamental SM fermions (bosons)
has a boson (fermion) super-partner. 
To reconcile the super-symmetry with experimental data, SUSY must be broken,
and 
sparticles are expected to be much heavier than their SM partners. Perhaps 
with an exception of the partner of the top quark~($t$), the stop quark, 
whose low mass eigenstate ($\tilde{t_1}$) could be actually lighter than the top quark.
%
It is interesting, that the mass inequality $m_{\tilde{t}_1} \lesssim m_t$ 
is demanded in the supersymmetric electroweak baryogenesis scenarios~\cite{EWB},
which attempt to provide an explanation for the origin of the baryon 
asymmetry in the Universe. 

If the chargino~($\chargino$) happens to be lighter than the stop quark, the decay 
$\tilde{t}_{1} \rightarrow b \chargino$ opens and becomes dominant.
At the Tevatron stop quarks, if exist, would be produced in pairs with the cross 
section a factor of $\sim 10$ smaller than that for the top quarks of the same mass.
Assuming that both charginos decay as $ \chargino \rightarrow  \neutralino \ell^{\pm} \nu$,
cascade decays of stop quarks  would produce experimental
event signatures with two high-$p_T$, oppositely charged leptons, 
two and more jets, and a large missing transverse energy $\met$. 
It is interesting, that this
event signature is identical to the dilepton final state of $\ttbar$ decays.
Therefore a small admixture of stop events to the $\ttbar$ dilepton events
would impact measurements of the properties of the top quark.
In particular, it  would bias the top mass measurements in the dilepton 
channel towards lower value relative to those measured 
in other channels of the $\ttbar$ decays. 
Intriguingly, it is exactly what had been observed in Tevatron experiments. 
The mass measurements in the lepton$+$jets and the dilepton 
channel based on the datasets corresponding to 1 fb$^{-1}$ of integrated 
luminosity  had only 7\% chance to yield results more discrepant than those
observed~\cite{mass}.
Using a conjecture above this feature could be interpreted 
as a sign of super-symmetry and it has served as a motivation for the following search.

\section{PRELIMINARY EVENT SELECTION AND STOP MASS RECONSTRUCTION}

We use data collected with the CDF detector
corresponding to 2.7~fb$^{-1}$ of integrated luminosity.
Events with a high-$p_T$ ($\geq$ 18 GeV/$c$) $e$ or
$\mu$ candidate are identified using the high speed trigger
electronics. In the preliminary event selection
stop candidate events are required to have two leptons ($e$ or
$\mu$), two or more jets with $E_T > 12 $ GeV
and missing transverse energy $\met > 20$ GeV. Events are also classified 
according to availability of a secondary vertex tag ($b$-tag)~\cite{secvtx}.
Optimized event selection criteria are determined 
at the last stage of the analysis.

The kinematic reconstruction of  stop events is a challenging task, 
since in each $\pairstop$ event there are only four particles 
(2 leptons and 2 $b$-jets), four-momenta of which are actually measured, 
while four other particles (2 $\nu$'s and 2 massive $\neutralino$'s)
escape detection, and their existence can only be 
inferred by an imbalance of transverse energy in the detector.
In addition, masses of $\chargino$ 
and $\neutralino$ are unknown, and therefore the kinematics of
stop events is severely unconstrained.
The $\pairstop$ event reconstruction is performed by constructing 
$\chi^2$-term, which represents the quadratic sum of the differences
between the true particle masses and invariant masses of their decay 
products as measured in the detector and 
divided by the respective width of the particle.
To overcome problems due to unconstrained kinematics, 
several useful simplifications are made.
First, we use $m_{\chargino}$ as a parameter in the fit and 
reconstruct stop mass in each event for several values of $m_{\chargino}$.
Second, to avoid the two-fold ambiguity in assigning a $b$-jet to the respective lepton,
we chose the pairing that yields the smallest sum of invariant masses $\sum m_{b\ell}$. 
This approach identifies the correct pairing in $\sim 90\%$ cases.
Third, the pair $\neutralino + \nu$ corresponding to each $\tilde{t}_1$ leg 
is treated as one massive particle with a large width. Our Monte Carlo 
studies showed that 
for a large range of neutralino masses $m_{\neutralino} \approx 46 - 90$ GeV, 
the choice of $m_{\neutralino + \nu}$ = 75 GeV and 
$\Gamma_{\neutralino + \nu}$ = 10 GeV works reasonably well.
Using all of these assumptions the sums of four-momenta $\neutralino + \nu$
corresponding to each stop leg are still not uniquely identified. Therefore
we integrate over the phase space of all possible solutions weighted 
by the $\chi^2$-term. 
The reconstructed mass of the stop quark is then given by
$m_{\tilde{t}_1}^{rec} = \int m_{\tilde{t}_1}^{rec, i} e^{-\chi_i^2} d S_i / \int  e^{-\chi_i^2} d S_i$
\section{BACKGROUNDS MODELING, SYSTEMATICS UNCERTAINTIES AND LIKELIHOOD FIT}

The dominant SM process that contributes to the dilepton $+$ jets event signature 
is $t\bar{t}$. Other SM processes include
 $Z/\gamma^* +$  jets, diboson production, and $W +$  
jet events, where one jet is misidentified as a lepton.
We use the \textsc{pythia}~\cite{PYTHIA} Monte Carlo (MC) event generator to simulate
$\pairstop$, $t\bar{t}$ and diboson processes.
$Z/\gamma^* +$ associated jet production is simulated with the \textsc{alpgen}~\cite{ALPGEN}
matrix element generator followed by parton fragmentation and hadronization by \textsc{pythia}. 
To model $W +$ jets events we exploit data events with one
fully identified lepton plus a lepton-like candidate that must fail
certain lepton ID requirements. A fake rate probability 
for such an object to be identified as a lepton is estimated from a large 
jet-triggered data sample, and then applied to each data lepton $+$ fake event.
We validate the modeling of dilepton events in the control regions
corresponding to low $\met$, zero and one jet bins and 
same-sign charged leptons. 

Imperfect knowledge of various experimental and theoretical 
parameters leads to systematic 
uncertainties which degrade our sensitivity to $\pairstop$ signal.
The dominant systematic effect is due to the uncertainties in the NLO 
theoretical cross sections for $\pairstop$ and $t\bar{t}$ production.
These uncertainties come from two sources: due to renormalization and factorization scale 
(11\% and 7\% for $\pairstop$ and $t\bar{t}$  respectively)
and due to parton density functions (14\% and 7\%)~\cite{stop,ttbar}.
We assume that the first one is uncorrelated 
between two sources, 
while the latter one is fully correlated.
The $t\bar{t}$ background is normalized to the theoretical cross section value
at the top mass world average of 172.5 GeV/c$^2$~\cite{top_mass} that is 
dominated by the measurements in the lepton $+$ jets channel of $\ttbar$ decays.

The experimental uncertainties include those due to jet energy scale~(3\%), 
$b$-tagging probability~(5\%), lepton ID and trigger efficiencies~(1\%), 
initial/final state radiation, and integrated luminosity~(6\%) which is applied to MC-based 
background estimates. Normalization of $W$ and $Z + $ jets events is determined 
from data. Uncertainty on $W$+jets is driven by the uncertainties in the fake rate 
predictions~(30\%), while uncertainty on $Z+$ jets is due to mis-modeling of the
high-$\met$ and jet multiplicity distributions and heavy-flavor corrections~(16\%). 

%

%


We employ the modified frequentist method, $CL_s$\cite{cls_descrip}, that represents
binned likelihood fits to the reconstructed stop mass 
of data
under the hypothesis of the SM background only and the hypothesis of signal plus background.
The fits are performed simultaneously in the channel with at least one $b$-tagged jet and 
the channel with no $b$-tags.
The systematic uncertainties for both signal and background, described above,
enter the fit as Gaussian constrained nuisance parameters.  
The shape uncertainties are accounted by allowing templates shapes to change 
(``morph'') according to the values of the nuisance parameters.  

The sensitivity of the likelihood fit (including all of the systematic uncertainties)
to the stop signal is tested for various event selection criteria
imposed separately for $b$-tagged and non-tagged channel. These criteria are allowed 
to vary using an algorithm based on biological evolution. Selection cuts yielding poor 
sensitivity to the signal are culled, while those improving sensitivity are bred together 
until reaching a plateau. Optimizing directly for the best 95\% C.L. limit has advantages  
with respect to event cuts selected based on some intermediate figure of merit.

%



%

\begin{table}[ht]
\hspace{-2cm}
  \begin{minipage}[b]{0.4\linewidth} \centering
   \includegraphics[width=8cm]{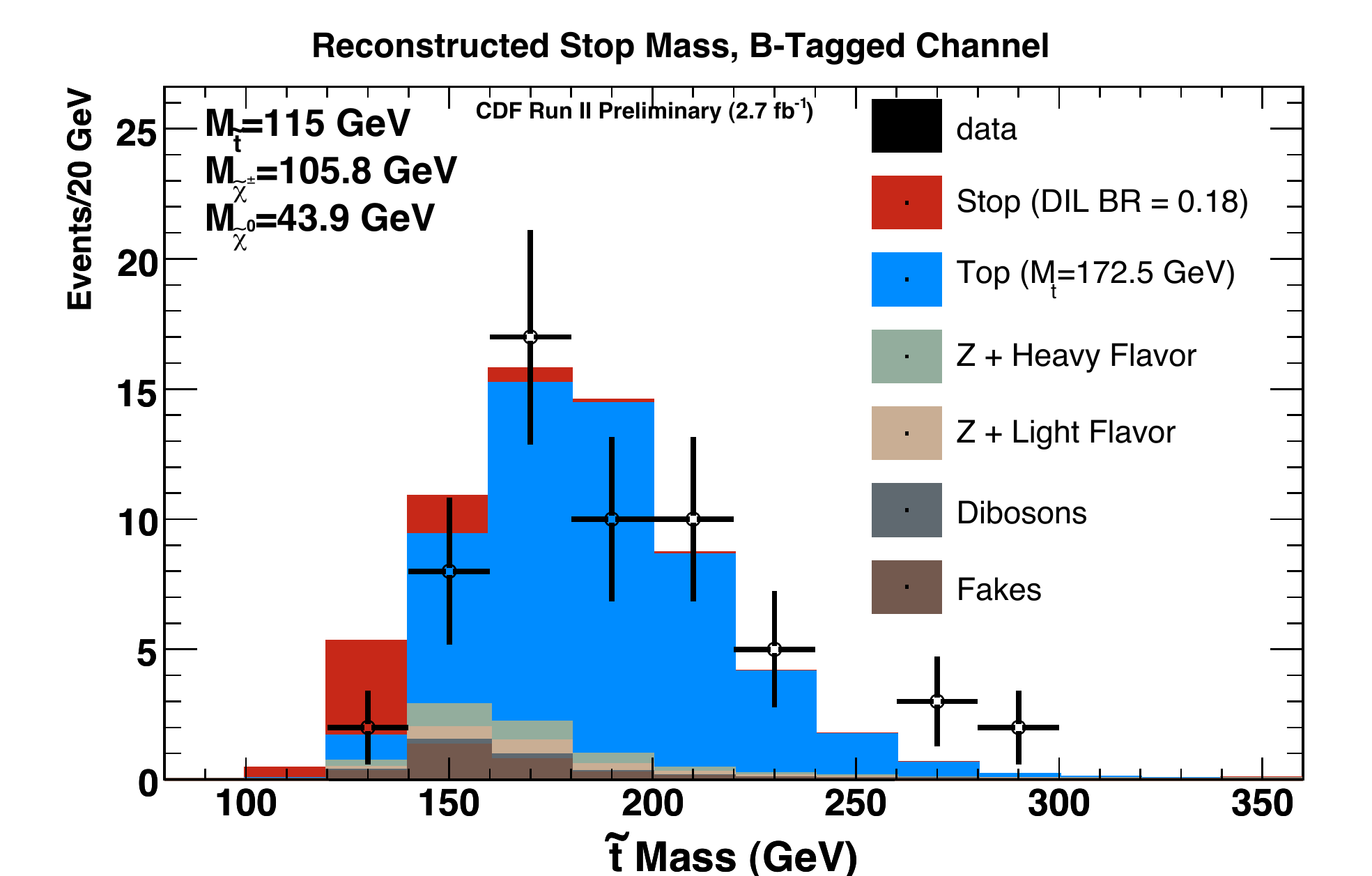}
\end{minipage}
 \hspace{1cm}
\begin{minipage}[b]{0.4\linewidth} \centering
  \begin{tabular}{c|cccc}
    \multicolumn{5}{c}{Events per 2.7 $fb^{-1}$ in the signal region with $\geq$ 1 $b$-tag.}
    \\
    \hline  
    Source                         &        $ee$             &     $\mu\mu$      &     $e\mu$           &       $\ell\ell$      \\
    \hline
    top                              &   11.3 $\pm$ 1.8 &   10.4 $\pm$ 1.6 &   26.7 $\pm$ 3.8 &   48.4 $\pm$ 7.0\\
    $Z/\gamma^*$+HF    &    1.3 $\pm$ 0.3  &    0.9 $\pm$ 0.2  &    0.4 $\pm$ 0.1  &    2.6 $\pm$ 0.5\\
    $Z/\gamma^*$+LF     &    0.9 $\pm$ 0.1  &    0.5 $\pm$ 0.1  &    0.3 $\pm$ 0.1  &    1.7 $\pm$ 0.1\\ 
    diboson                       &    0.2 $\pm$ 0.1  &    0.1 $\pm$ 0.1  &    0.3 $\pm$ 0.1  &    0.6 $\pm$ 0.1\\
    fake lepton                  &    0.5 $\pm$ 0.2  &    0.5 $\pm$ 0.1  &    1.8 $\pm$ 0.5  &    2.8 $\pm$ 0.8\\
    \hline
    Total                           &   14.2 $\pm$ 2.0 &   12.4 $\pm$ 1.6 &   29.4 $\pm$ 3.8 &   56.0 $\pm$ 7.3\\
    \hline
    stop                            &    1.1 $\pm$ 0.3  &    1.4 $\pm$ 0.4  &    3.0 $\pm$ 0.7  &    5.5 $\pm$ 1.2\\
    \hline
    \hline
    Data                           &      15                   &      12                   &      30                   &      57             \\
    \hline
       \end{tabular}  
   \label{tab:acc_table}
  \end{minipage}
 \caption{The reconstructed stop mass distribution (left) and the table of expected event yields (right) 
 		from Standard Model sources and $\pairstop$ at the 95\% C.L. exclusion limit(right).
              }
\end{table}

\begin{figure}[ht]
\includegraphics[width=8.5cm]{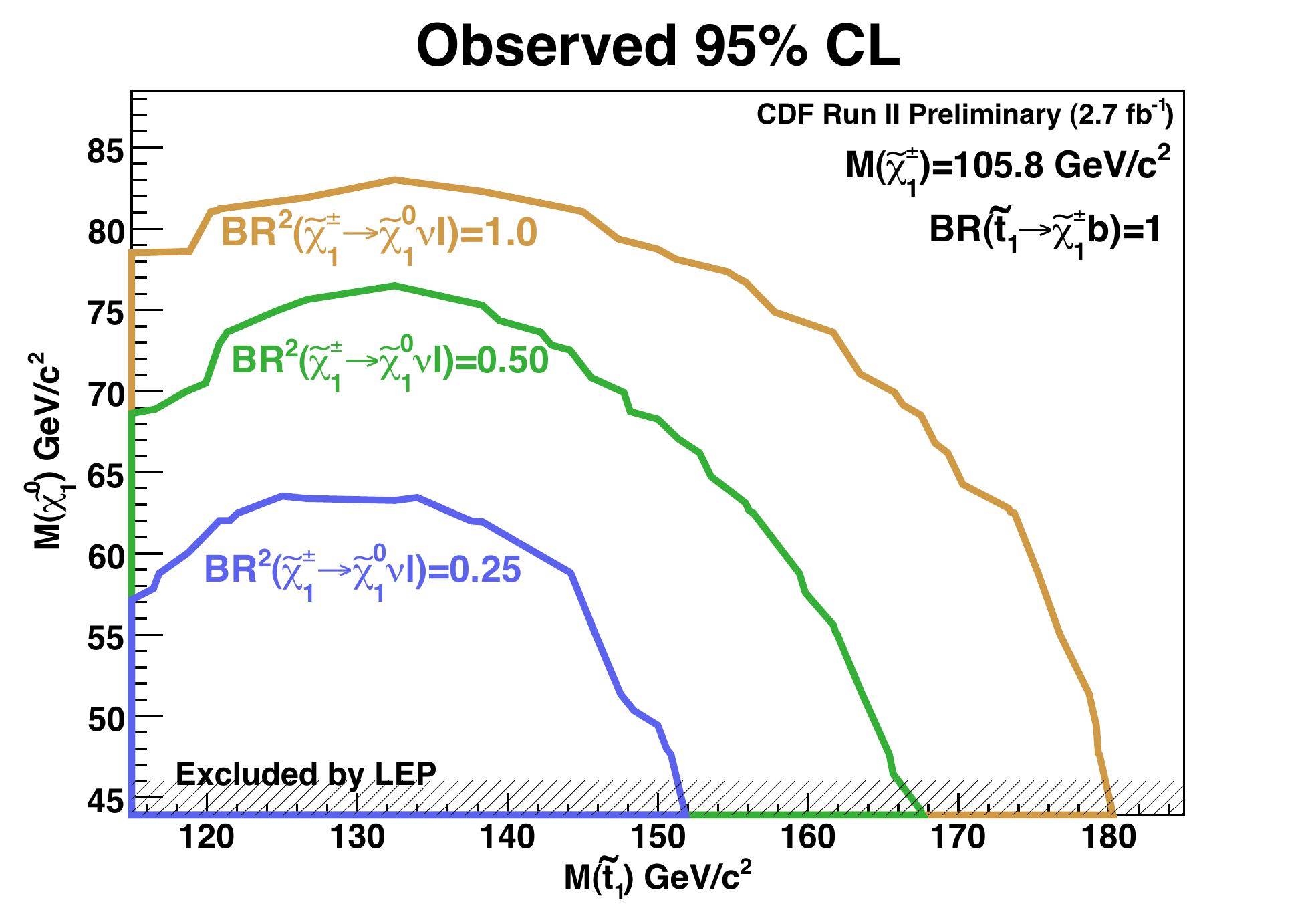}
\includegraphics[width=8.5cm]{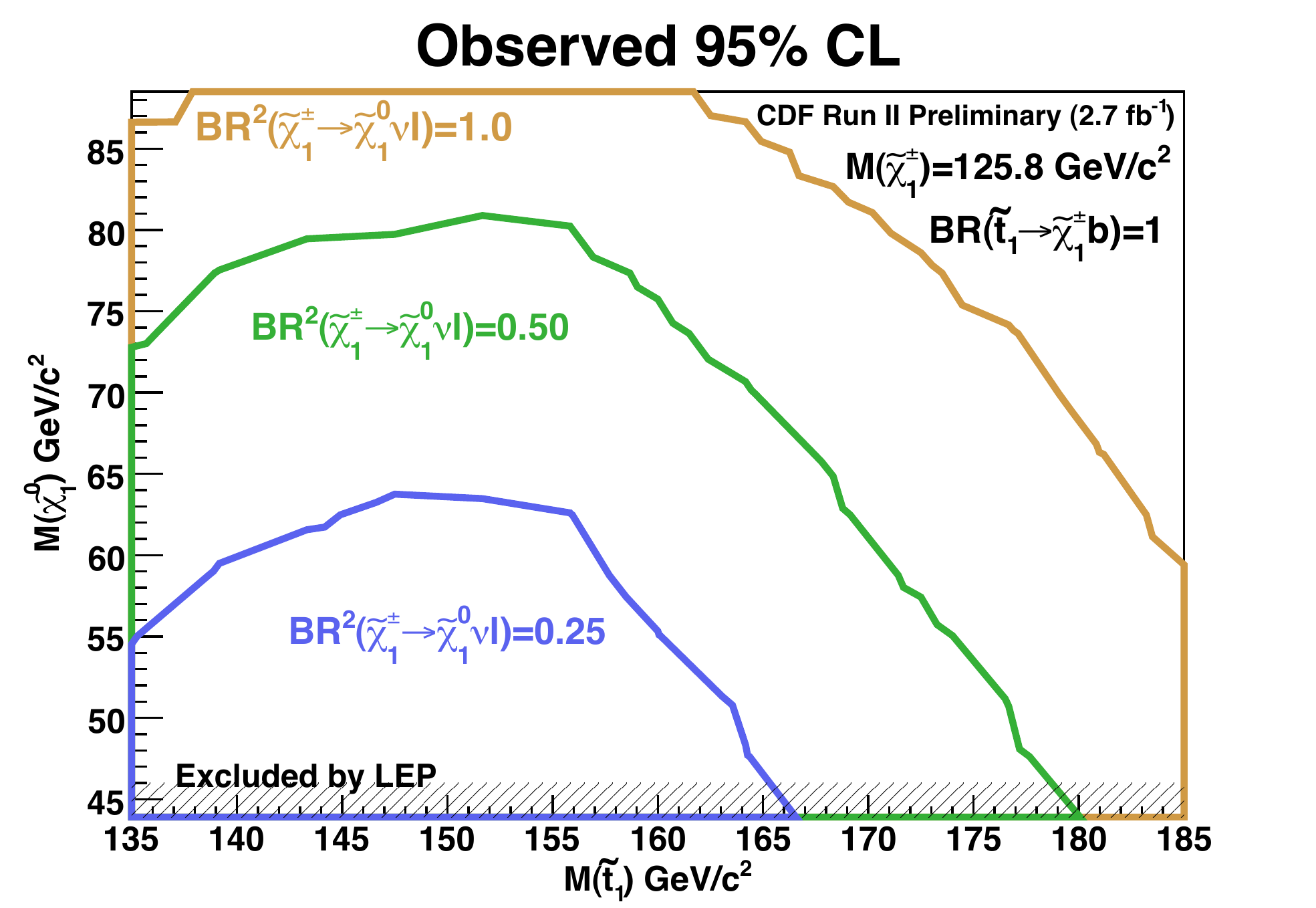}
\caption{ 
                 Observed 95\% C.L. limits on the 
                   $m_{\neutralino}$  and $m_{\tilde{t}_1}$  for several values of
                 ${\cal B}$($\chargino\rightarrow\neutralino\ell^{\pm}\nu$), and
                  	assuming $m_{\chargino} = $105.8 GeV/$c^2$ (left)
                  and $m_{\chargino} = $125.8 GeV/$c^2$ (right).
      		Universality of $e$, $\mu$, and $\tau$
                  in the $\chargino$ decays is assumed.
                }
\label{fig:Limits}
 \end{figure}

\section{RESULTS}

As can be seen from Table I, 
the data is consistent with the Standard Model.
The fit to the reconstructed stop mass 
distribution reveals no evidence of
$\pairstop$ production,
and thus we proceed and place 95\% CL limits on
$m_{\neutralino} $ and $m_{\tilde{t}_1}$ for several values 
of branching ratio  ${\cal B}(\chargino\rightarrow\neutralino\ell^{\pm}\nu)$
and $m_{\chargino}$. 
The results are presented in Figure~\ref{fig:Limits}.
%
%


\end{document}